\documentclass[9pt,twocolumn,twoside]{osajnl}

\usepackage{graphicx,subfigure}

\usepackage{multirow}

\usepackage{mathtools}

\DeclarePairedDelimiter{\ceil}{\lceil}{\rceil}

\journal{jocn} 

\setboolean{shortarticle}{false}

\title{Leveraging Multi-Step Traffic Forecasts for Multi-Period Planning Optical Networks}

\author[1,*]{Giannis Savva}
\author[1]{Hafsa Maryam}
\author[1]{Venkatesh Chebolu}
\author[1]{Tania Panayiotou}
\author[1]{Georgios Ellinas}

\affil[1]{KIOS Research and Innovation Center of Excellence, Department of Electrical and Computer Engineering, University of Cyprus, Nicosia, Cyprus}

\affil[*]{savva.giannis@ucy.ac.cy}

\begin{abstract}
In this work, multi-step traffic predictions are leveraged to enable multi-period planning in reconfigurable optical networks. The proposed framework aims to achieve spectrum savings by adapting the network to predicted time-varying conditions while ensuring the necessary quality-of-service (QoS) levels. Since frequent network (re)configurations may lead to undesired service disruptions, traffic predictions spanning various prediction horizons are exploited to balance the trade-off between spectrum savings and service disruptions. For multi-step-ahead prediction, an encoder-decoder deep learning model is employed to analyze real traffic traces. Subsequently, an Integer Linear Programming (ILP) formulation and heuristic algorithms are developed that use the predictions to proactively (re)optimize future network configurations, enhancing spectrum efficiency while minimizing service disruptions. The approaches are utilized under different scenarios, with the ILP achieving better solutions overall, and the heuristics achieving solutions close to the ILP at significantly lower running times. Further, the results present the effect of the prediction horizon on disruptions and over- and under- provisioning, showcasing that the prediction horizon selection greatly depends on the network operator targets in both network performance and predefined service level agreements.
\end{abstract}

\setboolean{displaycopyright}{false} 

\begin{document}

\maketitle

\section{Introduction}
With the emergence of advanced technologies such as 6G networks, the Internet of Things, and applications such as AI-generated content, immersive entertainment, smart cities, virtual reality, autonomous vehicles, and industrial automation, the demand for seamless and high-quality network connectivity has increased exponentially~\cite{ericsson2024, huawei2024}. Recent industry reports predict that global mobile data traffic is projected to double, reaching 303 exabytes per month by 2030~\cite{ericsson2024}, while only AI-driven data generation will generate over 1000 exabytes of data traffic by 2026~\cite{huawei2024}.

To cope with the growing demand, the use of traffic-driven machine learning (ML)-powered frameworks has become crucial for network optimization, enabling more effective planning and reducing energy consumption and operational costs~\cite{SurveyArxiv22,Natalino:24}. In general, traffic modeling enables multi-period planning, as traffic predictions can be used for proactive network (re)optimization to ensure efficient resource utilization, while meeting targeted QoS requirements in future network configurations~\cite{8620207}. 

While various ML models were exploited for network traffic prediction, those specifically designed to model sequential data, such as deep neural networks (DNNs) with gated recurrent units (GRU), long short-term memory (LSTM) units~\cite{Balanici:21,9782838, 10619909,pandaICC}, and DNNs with attention and self-attention mechanisms~\cite{LI2020102258, LU2025110913,QIN2024110674,10526615} were shown to exhibit competitive performance accuracy~\cite{SurveyArxiv22}, mainly due to their ability to efficiently capture the nonlinear nature and long-term temporal dependencies of network traffic. For predictive service provisioning, several optimal and heuristic-based algorithms have been developed to exploit the traffic predictions towards improving spectrum utilization~\cite{9748600, maryam2023uncertainty,Uzunidis:25}, service disruptions~\cite{6831425,10648008}, energy consumption~\cite{Alvizu2017EnergyED,10632183} and fairness in resource allocation decisions~\cite{9347908, 10214551, pandaICC,9322381}.

\subsection{Related Work}
The vast majority of existing work addresses \textit{traffic-driven service provisioning} by exploiting \textit{single-step-ahead predictions}~\cite{9748600, maryam2023uncertainty, Uzunidis:25, 6831425, 10648008, Alvizu2017EnergyED, 10632183, 8436062, 8501524, XIONG201999, 9203477, 8501527, 10526615}. For example, in~\cite{8436062}, a predictive resource allocation scheme in multi-domain elastic optical networks (EONs) is examined. In that scheme, intra-domain traffic forecasts obtained via DNNs were used by an online routing, modulation, and spectrum allocation (RMSA) heuristic to dynamically reconfigure intra-domain virtual links. The predictive scheme was shown to outperform a conventional RMSA approach (without predictions) in terms of connection blocking probability. Further, in~\cite{8501524} an online routing, spectrum, and core allocation (RSCA) algorithm was proposed for dynamic connection provisioning in data center multi-core EONs (DC-MC-EONs), leveraging single-step LSTM predictions. Forecasts were used to update, in real time, the mean residual lifetime of active connections, enabling better resource planning. This predictive RSCA method demonstrated superior performance compared to conventional RSCA, particularly when the LSTM model produced accurate predictions, highlighting the critical role of model accuracy. Similarly,~\cite{XIONG201999} presented a predictive RSCA heuristic using single-step Elman neural network forecasts to reduce spectrum fragmentation and inter-core crosstalk in MC-EONs. The proposed approach outperformed non-predictive dynamic schemes across metrics such as connection blocking, fragmentation, and spectrum utilization.

\begin{figure*}[t!]
\centering
{\includegraphics[scale =0.54]{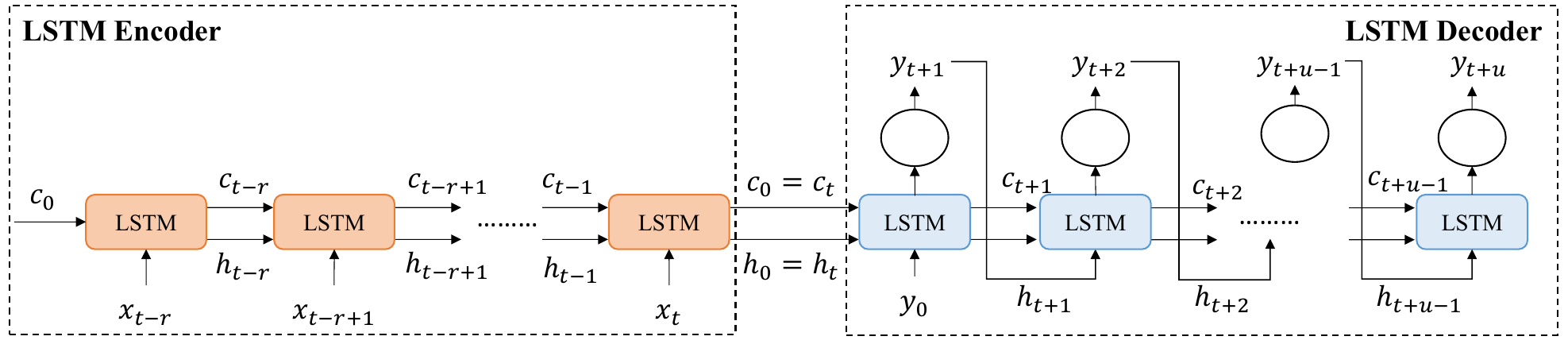} 
\caption{A generic ED-LSTM architecture.}
\label{lstm}}
\vspace{-0.1in}
\end{figure*}

In~\cite{9203477}, a graph convolutional generative (GCN-GAN) model and an LSTM model were used to produce hourly single-step link load predictions. Those forecasts were exploited by a dynamic RMSA heuristic. The GCN-GAN-guided version outperformed the LSTM-guided approach, mainly due to its higher prediction accuracy and ability to capture spatio-temporal traffic dependencies. Moreover, the authors in~\cite{8501527} leveraged single-step traffic predictions from Gaussian processes to support offline re-optimization in an IP-over-WDM network (i.e., multi-layer optimization). Several heuristics were proposed, each enabling re-optimization at different network layers. The predictive schemes outperformed static planning approaches in both spectrum utilization and regenerator cost savings. Additionally, coordinated multi-layer optimization was shown to provide greater benefits compared to optimizing each layer independently. Finally, in~\cite{10526615}, the authors demonstrated a real-time ML-assisted control loop for autonomous capacity provisioning in an optical metro-aggregation network testbed. A Temporal Fusion Transformer (TFT) model was employed for single-step traffic forecasting, supporting near real-time decision-making.

While multi-step-ahead traffic prediction (e.g., forecasting several hours into the future) holds significant potential, it remains largely underexplored. Although the general problem of multi-step forecasting was investigated in~\cite{Balanici:21, 9782838}, only~\cite{9782838} applied these predictions to service provisioning in optical networks. In that work, an encoder-decoder LSTM (ED-LSTM) model was employed to forecast future traffic demand matrices over multiple time steps in an optical data center and high-performance computing network. The resulting forecasts were then processed using an unsupervised clustering method to detect substantial changes in upcoming traffic patterns. A network reconfiguration was triggered whenever a forecasted demand matrix fell into a cluster different from that of the previous forecast. The study showed that having foresight into future traffic dynamics helped reduce service disruption and led to better end-to-end performance (specifically lower packet latency and loss) compared to static, non-adaptive network configurations.

\subsection{Contribution}
The focus of this work is to develop predictive routing and spectrum allocation (RSA) algorithms by exploiting multi-step-ahead predictions. An ED-LSTM architecture is initially employed, well suited to sequence-to-sequence tasks~\cite{cho-etal-2014-learning}, to enable multi-step-ahead predictions. These predictions are then exploited by the proposed multi-period planning algorithms aiming to minimize service disruptions and to ensure that spectrum resources are efficiently utilized. Benchmarking against RSA schemes that rely only on single-step predictions shows that the multi-step schemes perform significantly better in terms of service disruptions, with small impacts on over- and under-provisioning.

Preliminary results of this work were first presented in~\cite{10648008}. This work greatly extends~\cite{10648008} by:
\begin{itemize}
\item  Developing, apart from multi-period RSA heuristics, integer linear programming (ILP) algorithms, to optimally exploit the multi-step-ahead predictions. 
\item Training traffic prediction models that vary in the predicted steps ahead. 
\item Examining the impact of predicted steps on the RSA algorithms in terms of the overall quality of the allocation provided.
\item Comparing the performance of ILP and heuristic algorithms according to various metrics such as blocking rate, service disruptions, over-provisioning, under-provisioning, spectrum utilization, and execution times, and the impact that tuning some input parameters (e.g., different methods, weights, etc.) have on these metrics.
\end{itemize}

The rest of the paper is organized as follows. Section \ref{LSTM} details the multi-step-ahead traffic-prediction framework, including the dataset, model training, and evaluation procedure.
Section \ref{ilp} presents the ILP-based multi-step-ahead RSA algorithm, while heuristic alternatives are introduced in Section \ref{Heuristics}. Section \ref{evaluation} reports the network-level results (in terms of connection-blocking probability, service disruptions, over-provisioning and under-provisioning) and compares the proposed approaches. Finally, Section \ref{con} concludes the work and outlines directions for future work.

\section{Multi-Step-Ahead Traffic Prediction} \label{LSTM}
An ED-LSTM model \cite{cho-etal-2014-learning} is leveraged to learn how to accurately forecast multi-step-ahead future traffic demand (e.g., bit-rates) based on past and present traffic traces. Past and present traffic traces are given as vector ${\bf x}_t=[x_{t-r}, \cdots, x_{t-1},x_t]$ and future traffic values are given as vector ${\bf y}_t=[y_{t+1}, y_{t+2}, \cdots, y_{t+u}]$, where $r$ is the number of past observations, $t$ is the present time interval during which the current traffic demand $x_t$ is observed, and $u$ is the number of future traffic predictions (i.e., the number of future planning intervals over which predictions are made). 

A generic ED-LSTM architecture is illustrated in Fig.~\ref{lstm}. According to this architecture, the LSTM cell takes an input traffic sequence available at time interval $t$ (i.e., vector ${\bf x}_{t}$), and updates its hidden state $h_{t}$ according to the input and the previous hidden state. The LSTM encoder, reads each input ${\bf x}_{t}$ sequentially to update the cell and hidden states accordingly. After reading the end of each input sequence, the encoder summarizes this input sequence in the cell state and hidden state vectors ${\bf c}_t$ and ${\bf h}_t$, respectively, that are then given as input to the decoder component. The decoder is responsible to recursively predict the traffic sequence $y_{t+1},.., y_{t+u}$ according to $y_{t'}=g(h_{t'})$ $ \forall t'=t+1,...,t+u$, where $g(\cdot)$ is the activation function of the output layer. As the ED-LSTM is trained as a regressor, the decoder's output is connected with a fully connected dense layer.

The Adam algorithm is utilized for model optimization over dataset $D$=${\{{\bf x}_t, {\bf y}_t\}_{t=1}^{n}}$ to reduce the mean squared error (MSE) loss function, where $n$ denotes the observed traffic sequences. After training, the ED-LSTM model predicts future traffic values as $y_{t+u'} = f(x_{t-r},...., x_{t-1},x_{t},y_{t+1},y_{t+2},....,y_{t+u'-1})$, where $u'$ represents any future time interval and $u'\leq u$. Note that, even though the LSTM cells in Fig.~\ref{lstm} unfold in time, in practice, the encoder-LSTM cell processes $r$ traffic demand observations in sequence, while the decoder-LSTM cell sequentially predicts future traffic demands $u$. Note that the implementation used in this work is based on the openly available code in~\cite{code}.

\begin{figure*}[htpb]
 \begin{center}
\includegraphics[scale =0.29]{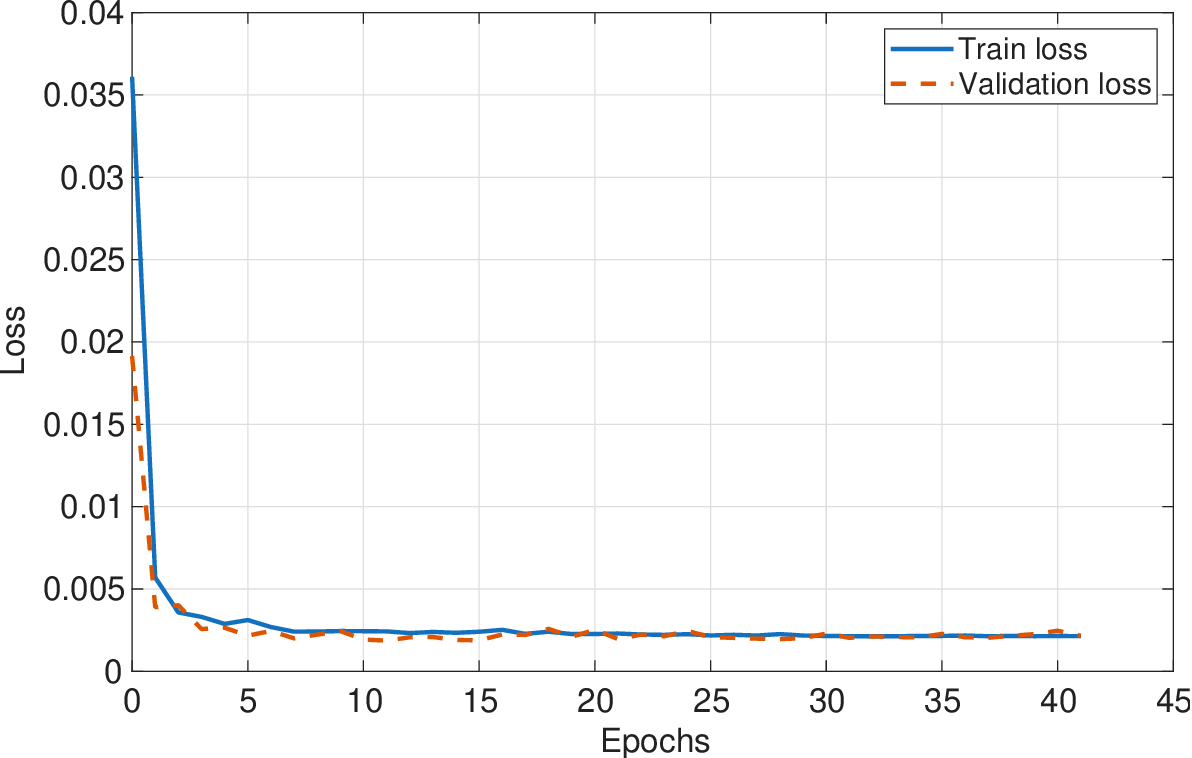}
\includegraphics[scale =0.29]{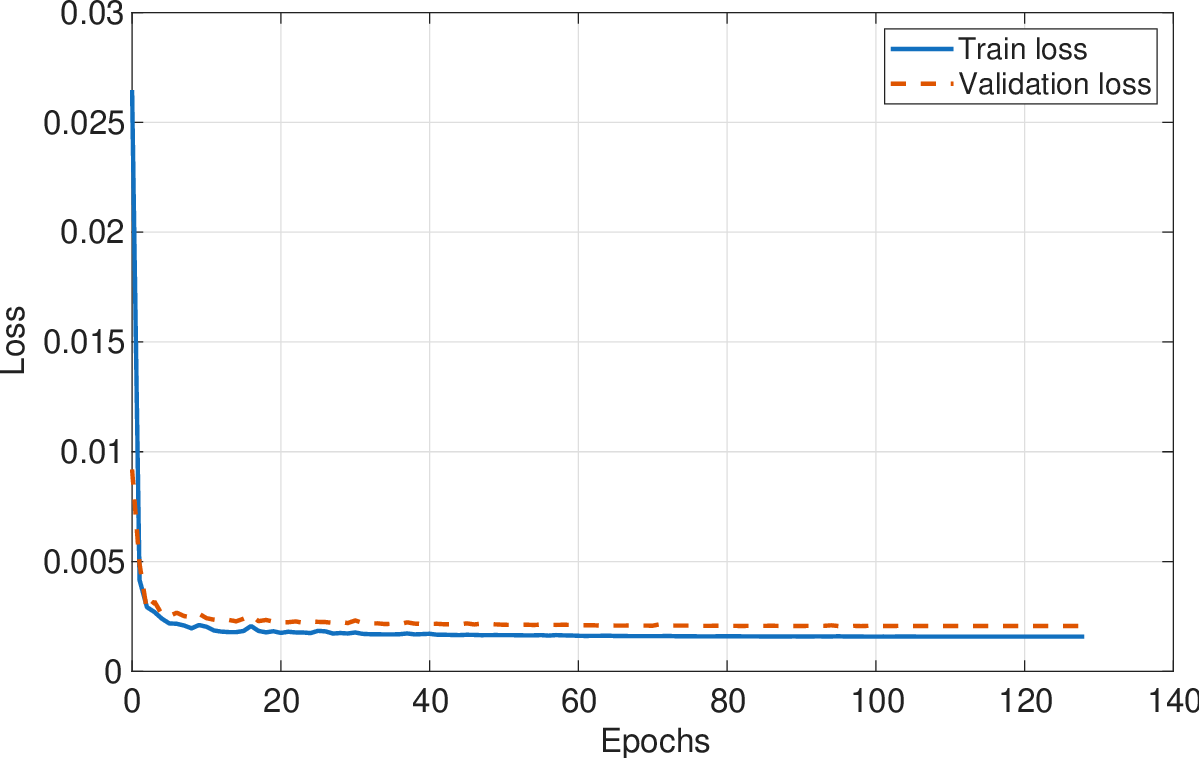}
\includegraphics[scale =0.29]{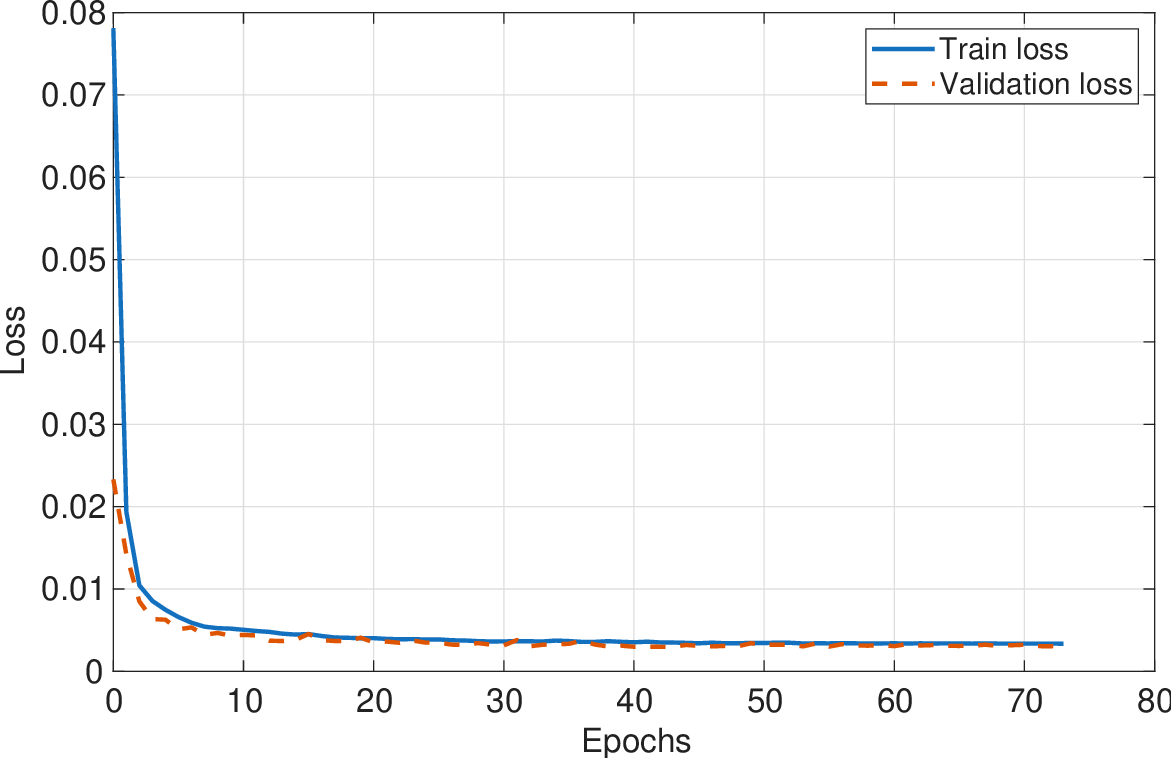}
\vspace{-0.1in}
 \caption{ED-LSTM training evolution vs. MSE loss for $D_1$ (node $8$ - $u$=$2$), $D_2$ (node $2$ - $u$=$4$), and $D_3$ (node $5$ - $u$=$6$) Abilene nodes.} 
\label{model}
\end{center}
\end{figure*} 

\begin{figure*}[htpb]
 \begin{center}
\includegraphics[scale =0.24]{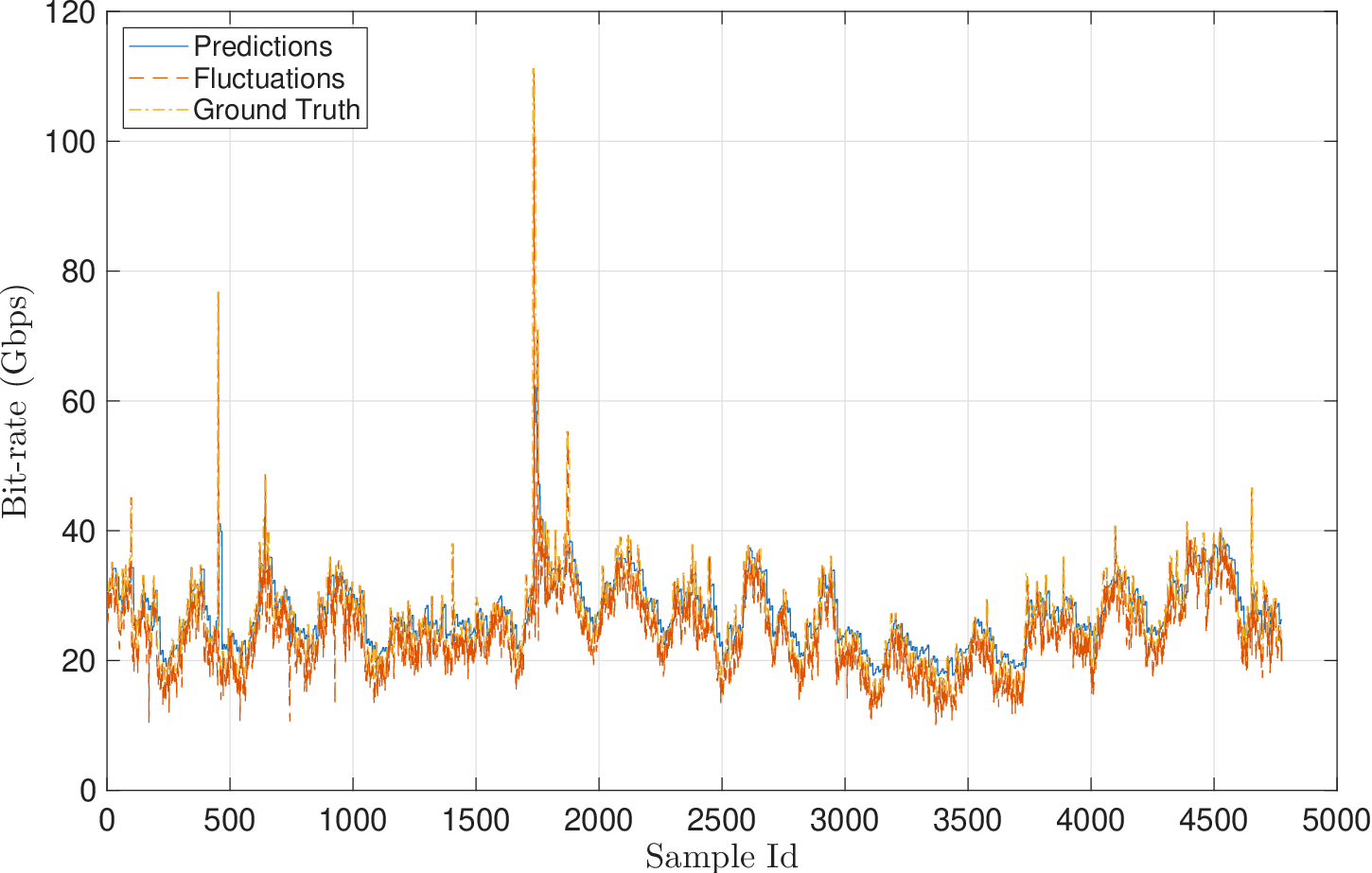}
\includegraphics[scale =0.24]{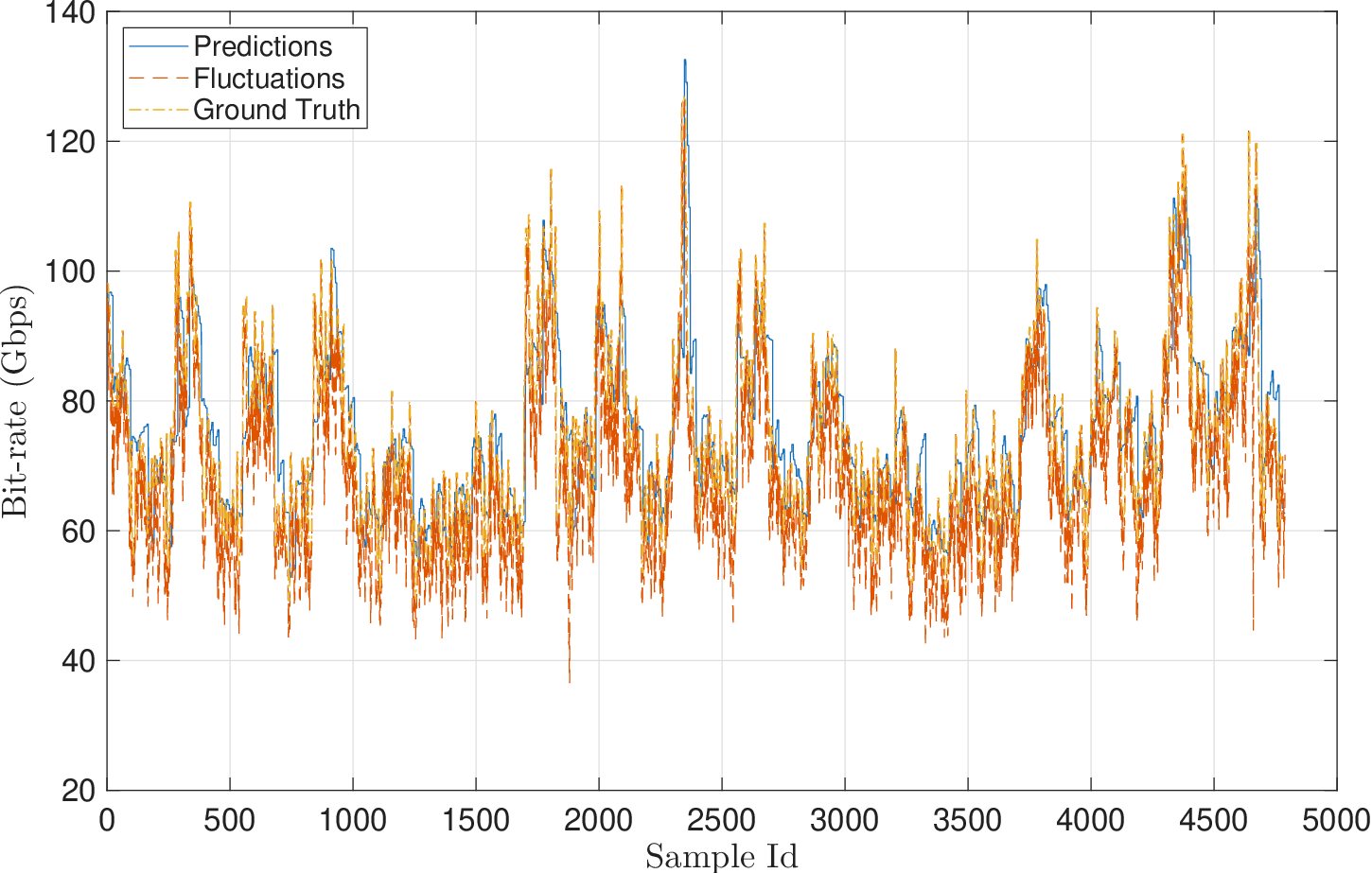}
\includegraphics[scale =0.24]{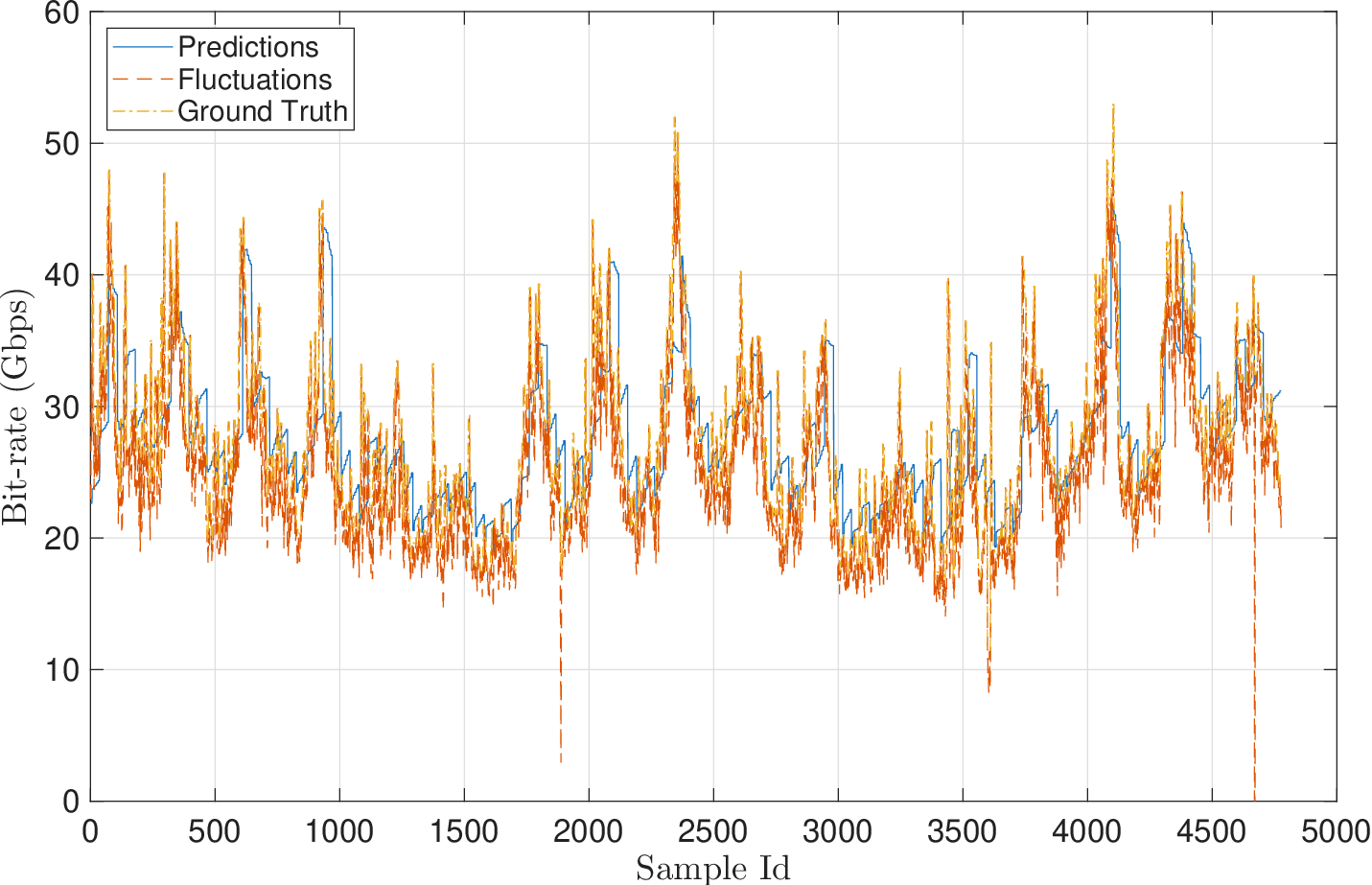}
 \vspace{-0.1in}
 \caption{Testing-Prediction Samples for $D_1$ (node $8$ - $u$=$2$), $D_2$ (node $2$ - $u$=$4$), and $D_3$ (node $5$ - $u$=$6$) Abilene nodes.} \label{model_testing}
\end{center}
\end{figure*} 

\subsection{Dataset Pre-processing} \label{dataset}
Dataset $D$ is created according to the real traffic traces of the $12$-node, $30$-link Abilene backbone network, which is openly available at http://sndlib.zib.de/home.action. This dataset, spanning a $6$-months period, provides bit-rate information (in Mbps) for every node pair in the form of traffic demand matrices and for every $5$ minutes. For simplicity, in this work, a dataset $D_v$ is created for each node $v$ in the network, with the traffic samples representing the aggregated bit-rate to node $v$ in $5$-minute intervals. As such, a different prediction model is trained for each source-destination pair $v$-$v'$. This is done to avoid possible model degradation (i.e., in achievable accuracy) which may occur when dealing with diverse sequence-to-sequence problems (i.e., with diverse time series~\cite{10252085}). Hence, input traffic patterns for each $D_v$ are created as ${\bf x}_t=[x_{t-r},..., x_{t}],\quad \text{where} \quad x_{t-i}=[\phi_{t-i}^1,...,\phi_{t-i}^k], \quad \forall i=0,1,...,r \, \forall t=1,2,...,T,$ where  
$t$ is the present planning interval, $r$ is the number of past planning intervals considered, and vector $x_t$ consists of $k$ traffic fluctuations in $t$. The ground-truths are created as ${\bf y}_t=[y_{t+1},..., y_{t+u}], \, \text{where} \, y_{t+j}= \max\{\phi_{t+j}^1,...,\phi_{t+j}^k\}, \forall j=0,1,..,u, \, \forall t=1,2..,T,$ 
targeting to predict the maximum traffic demand in $t$ with the aim of avoiding under-provisioning. Note that for the creation of each $D_v$=${\{{\bf x}_t, {\bf y}_t\}_{t=1}^{n}}$, the sliding window approach is used. 

Following the original dataset, where bit-rate information is provided according to a $5$-minute scale, $D_v$ datasets are created assuming that fluctuations occur on the same time scale. As time intervals spanning a period of $\tau$=$30$ minutes are assumed, the equivalent number of fluctuations that occur within a time interval will be $k$=$6$. In the rest of this work, prediction of traffic demands for the next $u$=$1,2,4,6$ time intervals are utilized. Note that the number of previous time intervals, $r$, is defined empirically (through various training examples), where for each multi-step-ahead value, $u$, the $r$ values that resulted to the best model accuracy are selected (as shown in Table \ref{hyper_acc_merged} and discussed below). 

\subsection{Model Training and Evaluation}
For each dataset $D_v$, a total of $T$=$4000$ consecutive traffic patterns are used for model training and evaluation. The data is normalized to fall within the range $[0, 1]$. Of these patterns, $80\%$ are allocated for training and validation, while the remaining $20\%$ are reserved for testing the model's accuracy.

Training is performed for various ED-LSTM models that differ in model configurations (i.e., hidden units, learning rate, batch size and $\delta$), as shown in Table \ref{test_cases_values} below: 

\begin{table}[htpb]
\centering
\caption{Different parameter values considered.}
\label{test_cases_values}
\begin{tabular}{|c|c|}
\hline
\textbf{Parameter} & \textbf{Values considered} \\ \hline
Sequence Length ($r$) & $[u+2, u+4, u+6]$ \\ \hline
Hidden Dimension & $[16,32,64,128]$ \\ \hline
Batch Size & $[64,128,256]$ \\ \hline
Learning Rate & $[0.1,0.01,0.001]$ \\ \hline
Delta ($\delta$) & $[0.5,1,1.5,2]$ \\ \hline
\end{tabular}
\end{table}


This is done to identify the best model configuration (i.e., the configuration leading to the most accurate model). As an early stopping criterion is set, model training continues either until model convergence or up to the number of predefined epochs (set up to $1000$). Further, a learning rate reduction is also applied with a factor of $0.5$ if no improvements are observed on validation loss over $10$ consecutive epochs. Table~\ref{hyper_acc_merged} summarizes the hyperparameters selected for each dataset and all prediction horizons utilized ($u=1,2,4,6$). The number of previous planning intervals, $r$, is also reported to highlight the fact that the length of past sequences is also an important parameter to be fine-tuned for accurately modeling multi-step-ahead traffic traces. 

\begin{table*}[t]
\centering
\scriptsize
\setlength{\tabcolsep}{3.7pt}
\caption{Selected hyperparameters and corresponding test accuracy for each dataset $D_v$ and prediction length $u$.}
\label{hyper_acc_merged}
\begin{tabular}{llcccccccccccc}
\toprule
\textbf{$u$} & \textbf{Metric}
& $D_1$ & $D_2$ & $D_3$ & $D_4$ & $D_5$ & $D_6$
& $D_7$ & $D_8$ & $D_9$ & $D_{10}$ & $D_{11}$ & $D_{12}$ \\
\midrule
\multirow{6}{*}{1}
& Sequence length
& 5 & 3 & 3 & 3 & 3 & 3 & 3 & 3 & 3 & 3 & 3 & 3 \\
& Hidden dimension
& 16 & 16 & 16 & 16 & 16 & 64 & 16 & 16 & 16 & 16 & 32 & 16 \\
& Batch size
& 128 & 256 & 64 & 64 & 64 & 64 & 128 & 64 & 64 & 64 & 64 & 64 \\
& Learning rate
& 0.100	 & 0.100 & 0.100 & 0.001 & 0.100 & 0.010 & 0.001 & 0.100 & 0.001 & 0.010 & 0.010 & 0.001 \\
& Delta
& 1.5 & 0.5 & 0.5 & 2 & 1.5 & 0.5 & 1 & 2 & 1 & 0.5 & 2	& 0.5 \\
& {Test accuracy}
& 0.009 & 0.0052 & 0.0016 & 0.0004 & 0.0015 & 0.0037 & 0.0123 & 0.0004 & 0.0065 & 0.012 & 0.023 & 0.002 \\
\midrule
\multirow{6}{*}{2}
& Sequence length
& 6 & 6 & 8 & 4 & 8 & 6 & 8 & 4 & 6 & 8 & 4 & 6 \\
& Hidden dimension
& 128 & 64 & 128 & 128 & 16 & 16 & 16 & 16 & 64 & 32 & 128 & 16 \\
& Batch size
& 128 & 256 & 64 & 256 & 128 & 128 & 64 & 64 & 64 & 64 & 128 & 256 \\
& Learning rate
& 0.01 & 0.001 & 0.01 & 0.01 & 0.1 & 0.01 & 0.01 & 0.01 & 0.01 & 0.01 & 0.01 & 0.1 \\
& Delta
& 1.5 & 0.5 & 1.5 & 2.0 & 0.5 & 1.5 & 0.5 & 0.5 & 1.5 & 1.5 & 0.5 & 0.5 \\
& Test accuracy
& 0.014 & 0.006 & 0.0024 & 0.0005 & 0.0021 & 0.0049 & 0.013 & 0.0005 & 0.0078 & 0.0157 & 0.0033 & 0.0024 \\
\midrule

\multirow{6}{*}{4}
& Sequence length
& 8 & 10 & 10 & 6 & 8 & 6 & 10 & 6 & 10 & 6 & 6 & 8 \\
& Hidden dimension
& 128 & 16 & 128 & 64 & 16 & 32 & 32 & 16 & 64 & 16 & 128 & 16 \\
& Batch size
& 128 & 64 & 128 & 256 & 64 & 128 & 128 & 64 & 128 & 64 & 128 & 256 \\
& Learning rate
& 0.01 & 0.001 & 0.01 & 0.01 & 0.1 & 0.001 & 0.01 & 0.1 & 0.01 & 0.1 & 0.01 & 0.1 \\
& Delta
& 2.0 & 0.5 & 0.5 & 0.5 & 0.5 & 2.0 & 2.0 & 1.5 & 1.0 & 1.5 & 1.5 & 0.5 \\
& Test accuracy
& 0.0197 & 0.0066 & 0.0037 & 0.0006 & 0.0029 & 0.0065 & 0.0151 & 0.0006 & 0.0093 & 0.0197 & 0.043 & 0.0029 \\
\midrule

\multirow{6}{*}{6}
& Sequence length
& 8 & 8 & 12 & 10 & 8 & 8 & 12 & 8 & 8 & 8 & 8 & 8 \\
& Hidden dimension
& 16 & 16 & 16 & 16 & 32 & 32 & 64 & 32 & 16 & 16 & 16 & 16 \\
& Batch size
& 256 & 128 & 64 & 256 & 128 & 128 & 64 & 64 & 64 & 128 & 64 & 64 \\
& Learning rate
& 0.01 & 0.1 & 0.01 & 0.01 & 0.01 & 0.001 & 0.01 & 0.01 & 0.01 & 0.1 & 0.1 & 0.1 \\
& Delta
& 1.0 & 0.5 & 1.0 & 0.5 & 1.0 & 0.5 & 0.5 & 1.5 & 2.0 & 1.5 & 1.0 & 1.5 \\
& Test accuracy
& 0.0219 & 0.0069 & 0.0043 & 0.0007 & 0.0036 & 0.0076 & 0.0157 & 0.0007 & 0.0104 & 0.0222 & 0.0049 & 0.0032 \\
\bottomrule
\end{tabular}
\end{table*}

Figure \ref{model} illustrates the evolution of the ED-LSTM model training of nodes ATLAM5, ATLAng, and CHINng for $u$=$2,4,6$, respectively, demonstrating that the loss function for all models is sufficiently optimized (i.e., converging to a sufficiently low test loss on their respective test dataset). A similar training behavior is observed for all datasets. Furthermore, Fig. \ref{model_testing} illustrates the test, prediction, and fluctuation values for the same nodes and datasets used in Fig. \ref{model}, demonstrating that predicted traffic closely follows the true traffic demand.  

Table \ref{hyper_acc_merged} provides the test accuracy for all $12$ datasets and for all the prediction steps, $u$, considered. Clearly, as the prediction horizon increases, model accuracy drops. As degraded model accuracy is expected to affect service level agreements (SLAs) in service provisioning (i.e., violations in targeted QoS) the prediction horizon must be carefully selected. Overall, in this work, models capable to predict the traffic demand with sufficient accuracy were successfully obtained; that is, accurate prediction of the traffic demand was achieved for up to $6$ time intervals ahead of time (i.e., for the next $3$ hours). This information is then exploited by both ILP and heuristic algorithms, which are presented in the section that follows.

\section{Multi-Period RSA: Proposed Approaches}~\label{ILP}


Given a graph $G(V,L)$, where $V$ represents the set of nodes and $L$ represents the set of edges, a planning horizon $u$, and a set of demands $C$, each associated with $u$ future predicted bit-rates, and the current network state (i.e., the current allocation of the connections in the network), the proposed approaches aim to provision connections in the network for the next $u$ time intervals with reduced service disruptions, while efficiently utilizing spectrum resources. It is also noted that for the following multi-period planning scenarios, the connection provisioning algorithm is activated every $u$ time intervals; that is, at every $t$, where $t\mod u$=$0$. For example, for $u$=$4$,  the RSA is solved at $t$=$0$ to accommodate the expected future traffic demand at $t$=$1,2,3,4$. Then, at $t$=$4$ the RSA is again solved again for considering future demands at $t$=$5,6,7,8$, and so on. 



For all algorithms, a pre-processing step takes place where a set of candidate paths (up to $\kappa$=$3$) are considered for each source-destination pair, calculated using Yen's Algorithm \cite{Yen_alg}. To account for the quality-of-transmission (QoT) of each connection when established using a path $p$, a conventional distance-adaptive modulation scheme is considered, where binary phase shift keying (BPSK), quadrature phase shift keying (QPSK), 8-quadrature amplitude modulation (8-QAM), and 16-quadrature amplitude modulation (16-QAM) modulation formats are utilized, based on the distance of the selected path $p$~\cite{5534599}. Hence, for each connection $c$, the bit-rate predicted for interval $t$ is converted into the number of frequency slots (FSs) required when using path $p$ via Eq.~\eqref{mod_eq}:

\begin{equation}
\label{mod_eq}
    \rho^{c}_{t,p}=\ceil{\frac{y^{c}_{t}}{B_{rate} \cdot MF_{p}}}
\end{equation}
where $y^{c}_{t}$ is the predicted bit-rate for connection $c$ at time interval $t$, $B_{rate}$ is the Baud rate (in Gbps), and $MF_{p}$ is the modulation format supported by candidate path $p$.

Further, as provisioning a connection for a next planning interval can potentially lead to connection reconfigurations (causing service disruptions and QoS violations), spectrum reduction-expansion~\cite{6381740} and connection re-allocation policies are considered for each connection, in an effort to avoid unnecessary service disruptions. Specifically, for each connection request, the following actions can be performed:
\begin{enumerate}
    \item \textit{Spectrum Reduction:} If the number of allocated FSs is greater than the FSs to be allocated for the next time intervals, then a spectrum reduction policy is applied (\textit{which is always feasible}). 
   \item \textit{Spectrum Expansion:} If the number of allocated FSs is less than the FSs to be allocated for the next time intervals, an in-place spectrum adjustment (i.e., allocating additional adjacent FSs on the same route) can be applied to minimize disruptions.
    \item \textit{Connection Re-allocation:} In case spectrum expansion is not feasible (i.e., additional adjacent FSs are not available), the connection is re-allocated (i.e., rerouted and/or reassigned new spectrum resources), leading to a disruption.
    \item \textit{Connection Blocking:} If re-allocation is not possible, then the connection is blocked.
\end{enumerate}

Further, in this work, it is assumed that a disruption occurs when a re-allocation is performed (i.e., the path utilized is changed, or a different central frequency must be allocated) for a connection.

Finally, it is noted that when $u$=$1$, all approaches are reduced to the conventional single-step-ahead prediction scheme, allocating to each connection request the bit-rate predicted for the next planning interval, which is used as a benchmark scenario.

\subsection{ILP Algorithm}\label{ilp}
The following section describes the proposed ILP formulation for solving the multi-period RSA problem regarding the next planning interval. In more detail, the proposed ILP considers the $u$ future predicted bit-rates of the next planning interval and the current network state, and provides as an output the RSA allocation of the connections in the network. 

\noindent \textbf{Parameters:}
\begin{itemize}\setlength\itemsep{0.06em}
    \item $G(V,L)$: A graph $G$, where $V$ is the set of nodes, and $L$ is the set of links;
    \item $\mathcal{T}$: Set of future time intervals, $|\mathcal{T}|=u$; 
    \item $C$: Set of connection requests;
    \item $c\in C$: A connection request;
    \item $F$: Set of frequency slots;
    \item $P^{c}$: Set of candidate paths for request $c$;
    \item $\rho^{c}_{i,p}$: Number of frequency slots requested by demand  $c$ in path $p$ for the time interval $t$;
    \item $\delta_{l,p}^{c}$: Equal to $1$ if the path $p$ of demand $c$ utilizes link $l$, and $0$ otherwise;
    \item $S_{f,p}^{c}$: Equal to $1$ if demand $c$ utilizes FS $f$ using path $p$ in the previous period, and $0$ otherwise;
    \item $\mathcal{R}^{p}_{c}$: Difference between the maximum and minimum number of FSs required by connection $c$ when using path $p$ [i.e., $\mathcal{R}^{p}_{c} = \max\limits_{i\in \mathcal{T}}{(\rho^{c}_{i,p})} - \min\limits_{i\in \mathcal{T}}{(\rho^{c}_{i,p}})$].
    \item $\mathcal{R}_{c}$: Maximum Difference between the maximum and minimum number of FSs required by connection $c$ over all candidate paths [i.e., $\mathcal{R}_{c} = \max\limits_{p\in P^{c}}{(\mathcal{R}^{p}_{c})}$].
    \item $w_{i},\; \forall i=1, ..., 5$: Objective function weight parameters.
\end{itemize}

\noindent \textbf{Variables:}
\begin{itemize}\setlength\itemsep{0.13em}
\item $q_{i,p}^{c}$: Binary, equal to $1$ if path $p$ and time interval $i$ are selected for demand $c$, and $0$ otherwise;
\item $R^{c}_{p}$: Binary, equal to $1$ if path $p$ is selected for demand $c$, and $0$ otherwise;
\item $W_{f,p}^{c}$: Binary, equal to $1$ if FS $f$ is allocated in path $p$ of demand $c$, and $0$ otherwise;
\item $\psi_{f,p}^{c}$: Binary, equal to $1$ if there exists a transition from free FS $(f-1)$ to FS $f$ that is allocated to demand $c$ in path $p$, and $0$ otherwise; 
\item $Y^{c}$: Binary, equal to $1$ if there is a disruption for demand $c$, and $0$ otherwise;
\item $Z^{c}$: Positive integer, associated with the difference between the required number of FS (based on predictions) and the allocated FSs of demand $c$;
\item $V^{c}$: Positive integer, associated with the difference between the allocated FSs and the required number of FS (based on predictions) for demand $c$;
\item $X_{l,f}$: Binary, equal to $1$ if slot $f$ in link $l$ is utilized, and $0$ otherwise.
\item $F_{max}$: Integer, equal to the maximum FS id utilized.
\end{itemize}

\noindent\textbf{Objective:}

\noindent \textit{Minimize:}
\begin{gather}
 \frac{w_{1}}{|C|} \cdot \sum_{c \in C} Y^{c} + \frac{w_{2}}{ 1+\sum\limits_{c\in C} \mathcal{R}_{c}} \cdot \sum_{c\in C} Z^{c} + \frac{w_{3}}{1+ \sum\limits_{c\in C} \mathcal{R}_{c}} \cdot \sum_{c\in C} V^{c} \nonumber\\ + \frac{w_{4}}{|L| \cdot |F|} \cdot \sum_{l\in L} \sum_{f\in F} X_{l,f} + \frac{w_{5}}{|F|} \cdot F_{max}
\end{gather}
The aim of the objective is to minimize the number of disruptions (first term), the difference between the allocated FS and the lowest (second term) and highest (third term) number of FS predicted (based also on the selected path), the slot utilization (fourth term), and the maximum slot id utilized (i.e., minimize the fragmentation - fifth term). Further, all terms are normalized in the objective (i.e., a ``$+1$'' is added in the denominator of the second and third terms to avoid division by $0$ in case $\sum\limits_{c\in C}\mathcal{R}_{c}=0$), while each term $i$ is also weighted by a parameter $w_{i}$.

\noindent \textit{Subject to the following constraints:}
\begin{gather}
	\label{constraint1}
	\sum_{p \in P^{c}} \sum_{i \in \mathcal{T}} q^{c}_{i,p} = 1\, ; \, \forall c \in C
\\
	\label{constraint2}
	\sum_{i \in \mathcal{T}} q^{c}_{i,p} = R^{c}_{p}\, ; \, \forall c \in C, \forall p \in P^{c}
\\
	\label{constraint3}
	\sum_{f \in F} W_{f,p}^{c} = \sum_{i \in \mathcal{T}} \rho^{c}_{i,p} \cdot q^{c}_{i,p} \, ; \, \forall p \in P^{c}, \forall c \in C
\\
\label{constraint4}
 \psi_{f,p}^{c} \geq
 \begin{cases}
  W_{f,p}^{c} - W_{f-1,p}^{c}  & \text{if} \, f>1 \\
  W_{f,p}^{c} & \text{if} \, f=1 \\
 \end{cases}; 
 \begin{cases}
\forall  p\in P^{c}, \\ \forall f \in F, \\ \forall c \in C
 \end{cases} 
\\
\label{constraint5}
\sum_{f \in F} {\sum_{p \in P^{c}}} \psi_{f,p}^{c} \leq 1 ; \quad \forall c \in C
\\
\label{constraint6}
\sum_{c \in C} \sum_{p \in P^{c}} W_{f,p}^{c} \cdot \delta_{l,p}^{c} \leq X_{l,f} ; \,\, \forall f \in F,\forall l \in L 
\\ 
f \cdot X_{l,f} \leq F_{max} ; \quad  \forall l \in L, \forall f \in F \label{constraint7}
\\
Y^{c}=1-\sum_{f\in F}\sum_{p\in P^{c}}  \psi^{c}_{f,p} \cdot S^{c}_{f,p}; \,\, \forall c \in C \label{constraint8}
\\
    Z^{c} \geq  \rho^{c}_{i,p}\cdot R^{c}_{p}- \sum_{f \in F} W_{f,p}^{c}; \begin{cases}
\forall i \in \mathcal{T}, \forall p \in P^{c}, \\ \forall c \in C
 \end{cases}\label{constraint9}
 \\
    V^{c} \geq \sum_{f \in F} W_{f,p}^{c} - \rho^{c}_{i,p}\cdot R^{c}_{p};\begin{cases}
\forall i \in \mathcal{T}, \forall p \in P^{c}, \\ \forall c \in C
 \end{cases}\label{constraint10} 
\end{gather}

Constraint (\ref{constraint1}) ensures that one candidate path (among the candidate path space) and a specific bit-rate requirement (among the predicted values for the time interval) is selected for each connection request, with Constraint (\ref{constraint2}) utilized to indicate the path selected for each connection. Further, Constraint (\ref{constraint3}) ensures that the number of slots allocated to a connection $c$ satisfies the bit-rate requirement selected for that connection, while Constraints (\ref{constraint4})-(\ref{constraint5}) are used to ensure the slot contiguity constraint, and Constraint (\ref{constraint6}) ensures the slot non-overlapping constraint. In addition, Constraint (\ref{constraint7}) is used to determine the maximum frequency slot id utilized ($F_{max}$), while Constraint (\ref{constraint8}) ascertains whether a connection $c$ is disrupted. It is important to note here that variable $Y^{c}=1$ only if the connection is re-allocated (i.e., disrupted), while $Y^{c}=0$ if spectrum expansion/reduction is selected by the ILP. Further, the difference between the maximum number of predicted and allocated FS is considered via Constraint (\ref{constraint9}), while the difference between the minimum number of predicted and allocated FS is considered via Constraint (\ref{constraint10}), as a means to relate over- and under-provisioning metrics within the objective of the ILP. It is important to note here that the slot continuity constraint  is implicitly taken into account by the definition of variable $W^{c}_{fp}$ (i.e., if $W^{c}_{fp}=1$, then the same slot $f$ is utilized across all links of path $p$ for connection $c$). 


The number of variables and constraints utilized for this ILP are provided below, in terms of the number of demands ($|C|$), number of FSs ($|F|$), number of candidate paths ($\kappa$), number of links ($|L|$), and number future time intervals ($|\mathcal{T}|=u$).

\begin{table}[htpb]
\small
\centering
\caption{Proposed ILP - Number of Variables and Constraints.}
\begin{tabular}{|c|c|}
\hline
\begin{tabular}[c]{@{}c@{}}Number of\\  Variables\end{tabular} & \begin{tabular}[c]{@{}c@{}} $|C|\cdot \kappa \cdot |\mathcal{T}|+2\cdot |C|\cdot \kappa \cdot |F|+$ \\ $+|C|\cdot \kappa+3\cdot |C| + |L|\cdot |F| + 1$ \end{tabular}  \\ \hline
\begin{tabular}[c]{@{}c@{}}Number of\\ Constraints\end{tabular} & \begin{tabular}[c]{@{}c@{}} $3\cdot |C|+2\cdot|C|\cdot \kappa +|C|\cdot \kappa \cdot |F| + $ \\ $ +2\cdot |L|\cdot |F| + 2\cdot |C|\cdot \kappa \cdot |\mathcal{T}| $ \end{tabular} \\ \hline
\end{tabular}
\end{table}

\subsection{Heuristic Algorithms}~\label{Heuristics}
\begin{figure}[htpb]
 \begin{center}
\includegraphics[scale =0.4]{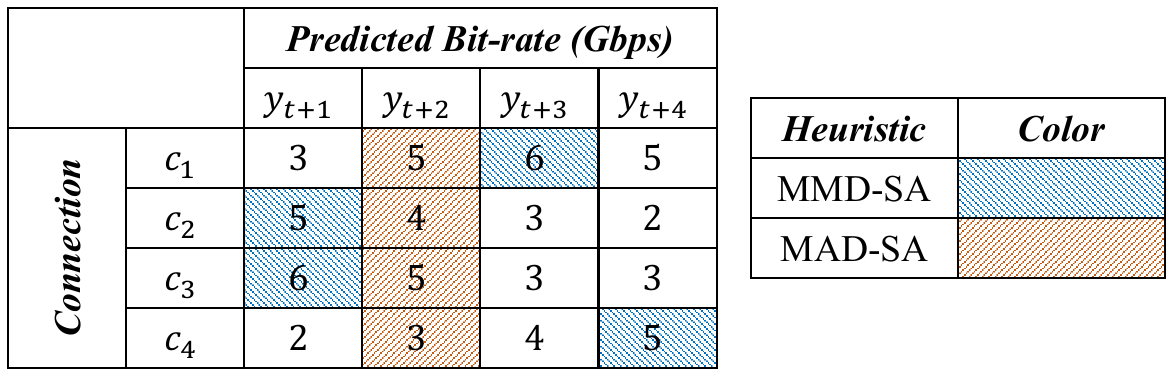}
 \caption{MMD-SA and MAD-SA examples given the bit-rates predicted for each time interval $t+i$ and for each connection $c_i$, with $u$=$4$.}
 \label{alg_example}
\end{center}
\end{figure}

The following spectrum allocation (SA) heuristic algorithms are proposed: i) the multi-step maximum demand SA (MMD-SA), and ii) the multi-step average demand SA (MAD-SA) heuristics. Both SA heuristics first solve the routing problem in a first-fit manner (considering the candidate paths with $\kappa=3$, sorted in ascending form based on the overall distance of each path), followed by a first-fit SA strategy. Hence, the SA schemes differ in the way the selected spectrum size is chosen for each connection request and future planning interval. Note that since a prediction interval $t+i$ is equivalent to a planning interval $t+i$ (i.e., an interval where RSA may be applied), these two terms are used interchangeably. To this end, heuristic descriptions follow:
\begin{itemize}
\item {\bf  MMD-SA}: For each connection request, the maximum bit-rate predicted across all $u$ future planning intervals is selected. An example is given in Fig.~\ref{alg_example} where for each connection, the maximum predicted bit-rate is indicated (i.e., 6,5,6,5 for connections $c_1,c_2,c_3,c_4$, respectively).

\item {\bf MAD-SA}: Each connection request is allocated to the bit-rate predicted for the $t+i$ future planning interval, where $t+i,~i\in {1,..,u}$ is the planning interval for which the aggregated predicted bit-rate among all connections is the maximum (i.e., $\max \{\sum_c y^c_{t+i}\}_{i=1}^u$, where $y^c_{t+i}$ is the bit-rate predicted at $t$ for the connection $c$ and for planning interval $t+i$). An example is given in Fig.~\ref{alg_example}, indicating the bit-rate to be allocated for each connection according to MAD-SA (i.e., 5,4,5,3 for connections $c_1,c_2,c_3,c_4$, respectively). In this example, $t+2$ is the planning interval for which the aggregated predicted bit-rate is the maximum (i.e., 17 Gbps). Hence, for each connection the bit-rates predicted for $t+2$ are selected to serve the next $u$ time intervals. 
\end{itemize}

For both heuristics, the selected bit-rates are chosen to accommodate the traffic demand for all connections and for all $u$ future time intervals, along with the path (among the precomputed $\kappa$-shortest paths) that can support each connection. The selected bit-rates are converted into the required number of frequency slots, following Eq. \eqref{mod_eq}. Also, in both heuristics, the demands to be established in the network are sorted in descending order, based on their bit-rates selected at each planning interval (i.e., each time the algorithm runs, it aims at establishing first connections that request higher bit-rates).

Further, in order for the heuristics to reduce unnecessary service disruptions, \textit{Spectrum Reduction} or \textit{Spectrum Expansion} is first applied (depending on the difference between the previously allocated FSs and the newly provided ones), before {\it Connection Re-allocation} is performed. Finally, if re-allocation is not possible, then the connection is blocked.

\begin{table*}[h]
\scriptsize
\setlength{\tabcolsep}{1.5pt}
\centering

\caption{Evaluation of service provisioning schemes: ILP and heuristic approaches.}
\label{ILP_heuristic_results}
\begin{tabular}{|c|cccc||cccc||cccc||cccc|}
\hline
\textbf{Approach} &
  \multicolumn{4}{c||}{ILP - $Sc_{1}:[20,20,1,0.01,10]$} &
  \multicolumn{4}{c||}{ILP - $Sc_{2}:[20,2,5,0.01,10]$} &
  \multicolumn{4}{c||}{Heuristic - MMD-SA} &
  \multicolumn{4}{c|}{Heuristic - MAD-SA} \\ \hline
\textbf{$u$} &
  \multicolumn{1}{c|}{\textbf{$1$}} &
  \multicolumn{1}{c|}{\textbf{$2$}} &
  \multicolumn{1}{c|}{\textbf{$4$}} &
  \textbf{$6$} &
  \multicolumn{1}{c|}{\textbf{$1$}} &
  \multicolumn{1}{c|}{\textbf{$2$}} &
  \multicolumn{1}{c|}{\textbf{$4$}} &
  \textbf{$6$} &
  \multicolumn{1}{c|}{$1$} &
  \multicolumn{1}{c|}{$2$} &
  \multicolumn{1}{c|}{$4$} &
  $6$ &
  \multicolumn{1}{c|}{$1$} &
  \multicolumn{1}{c|}{$2$} &
  \multicolumn{1}{c|}{$4$} &
  $6$ \\ \hline
\textbf{Blocking} &
  \multicolumn{1}{c|}{$0$} &
  \multicolumn{1}{c|}{$0$} &
  \multicolumn{1}{c|}{$0$} &
  $0$ &
  \multicolumn{1}{c|}{$0$} &
  \multicolumn{1}{c|}{$0$} &
  \multicolumn{1}{c|}{$0$} &
  $0$ &
  \multicolumn{1}{c|}{$0$} &
  \multicolumn{1}{c|}{$0$} &
  \multicolumn{1}{c|}{$0$} &
  $0$ &
  \multicolumn{1}{c|}{$0$} &
  \multicolumn{1}{c|}{$0$} &
  \multicolumn{1}{c|}{$0$} &
  $0$ \\ \hline
\textbf{Disruptions} &
  \multicolumn{1}{c|}{$6$} &
  \multicolumn{1}{c|}{$0$} &
  \multicolumn{1}{c|}{$0$} &
  $0$ &
  \multicolumn{1}{c|}{$6$} &
  \multicolumn{1}{c|}{$1$} &
  \multicolumn{1}{c|}{$0$} &
  $0$ &
  \multicolumn{1}{c|}{$6$} &
  \multicolumn{1}{c|}{$5$} &
  \multicolumn{1}{c|}{$3$} &
  $3$ &
  \multicolumn{1}{c|}{$6$} &
  \multicolumn{1}{c|}{$5$} &
  \multicolumn{1}{c|}{$3$} &
  $3$ \\ \hline
\textbf{Under-provisioning (\textit{Gbps})} &
  \multicolumn{1}{c|}{$2.72$} &
  \multicolumn{1}{c|}{$2.97$} &
  \multicolumn{1}{c|}{$3.45$} &
  $3.88$ &
  \multicolumn{1}{c|}{$2.72$} &
  \multicolumn{1}{c|}{$3.32$} &
  \multicolumn{1}{c|}{$4.25$} &
  $4.93$ &
  \multicolumn{1}{c|}{$2.72$} &
  \multicolumn{1}{c|}{$2.97$} &
  \multicolumn{1}{c|}{$3.45$} &
  $3.88$ &
  \multicolumn{1}{c|}{$2.72$} &
  \multicolumn{1}{c|}{$3.02$} &
  \multicolumn{1}{c|}{$3.55$} &
  $4.04$ \\ \hline
\textbf{Over-provisioning (\textit{Gbps})} &
  \multicolumn{1}{c|}{$5.93$} &
  \multicolumn{1}{c|}{$8.01$} &
  \multicolumn{1}{c|}{$9.46$} &
  $10.41$ &
  \multicolumn{1}{c|}{$5.93$} &
  \multicolumn{1}{c|}{$6.24$} &
  \multicolumn{1}{c|}{$5.58$} &
  $5.43$ &
  \multicolumn{1}{c|}{$5.93$} &
  \multicolumn{1}{c|}{$8.01$} &
  \multicolumn{1}{c|}{$9.46$} &
  $10.41$ &
  \multicolumn{1}{c|}{$5.93$} &
  \multicolumn{1}{c|}{$7.75$} &
  \multicolumn{1}{c|}{$8.64$} &
  $9.74$ \\ \hline
\textbf{Under-provisioning (\textit{FS})} &
  \multicolumn{1}{c|}{$0.13$} &
  \multicolumn{1}{c|}{$0.15$} &
  \multicolumn{1}{c|}{$0.18$} &
  $0.19$ &
  \multicolumn{1}{c|}{$0.13$} &
  \multicolumn{1}{c|}{$0.17$} &
  \multicolumn{1}{c|}{$0.22$} &
  $0.26$ &
  \multicolumn{1}{c|}{$0.13$} &
  \multicolumn{1}{c|}{$0.15$} &
  \multicolumn{1}{c|}{$0.18$} &
  $0.19$ &
  \multicolumn{1}{c|}{$0.13$} &
  \multicolumn{1}{c|}{$0.15$} &
  \multicolumn{1}{c|}{$0.18$} &
  $0.20$ \\ \hline
\textbf{Over-provisioning (\textit{FS})} &
  \multicolumn{1}{c|}{$0.45$} &
  \multicolumn{1}{c|}{$0.56$} &
  \multicolumn{1}{c|}{$0.60$} &
  $0.68$ &
  \multicolumn{1}{c|}{$0.45$} &
  \multicolumn{1}{c|}{$0.47$} &
  \multicolumn{1}{c|}{$0.42$} &
  $0.41$ &
  \multicolumn{1}{c|}{$0.45$} &
  \multicolumn{1}{c|}{$0.56$} &
  \multicolumn{1}{c|}{$0.60$} &
  $0.68$ &
  \multicolumn{1}{c|}{$0.45$} &
  \multicolumn{1}{c|}{$0.54$} &
  \multicolumn{1}{c|}{$0.56$} &
  $0.63$ \\ \hline
\textbf{Spectrum Utilization (\textit{FS})} &
  \multicolumn{1}{c|}{$150.96$} &
  \multicolumn{1}{c|}{$153.82$} &
  \multicolumn{1}{c|}{$154.19$} &
  $157.36$ &
  \multicolumn{1}{c|}{$150.96$} &
  \multicolumn{1}{c|}{$150.54$} &
  \multicolumn{1}{c|}{$147.60$} &
  $146.21$ &
  \multicolumn{1}{c|}{$159.59$} &
  \multicolumn{1}{c|}{$162.91$} &
  \multicolumn{1}{c|}{$163.14$} &
  $165.67$ &
  \multicolumn{1}{c|}{$159.59$} &
  \multicolumn{1}{c|}{$161.60$} &
  \multicolumn{1}{c|}{$161.21$} &
  $163.37$ \\ \hline
\textbf{$F_{max}$} &
  \multicolumn{1}{c|}{$70.94$} &
  \multicolumn{1}{c|}{$44.25$} &
  \multicolumn{1}{c|}{$34.16$} &
  $42.34$ &
  \multicolumn{1}{c|}{$70.94$} &
  \multicolumn{1}{c|}{$32.17$} &
  \multicolumn{1}{c|}{$51.19$} &
  $35.68$ &
  \multicolumn{1}{c|}{$42.35$} &
  \multicolumn{1}{c|}{$40.36$} &
  \multicolumn{1}{c|}{$25.56$} &
  $38.02$ &
  \multicolumn{1}{c|}{$42.35$} &
  \multicolumn{1}{c|}{$40.28$} &
  \multicolumn{1}{c|}{$25.31$} &
  $37.98$ \\ \hline
\textbf{Running Time} &
  \multicolumn{1}{c|}{$2.55\;s$} &
  \multicolumn{1}{c|}{$3.59\;s$} &
  \multicolumn{1}{c|}{$4.86\;s$} &
  $5.96\;s$ &
  \multicolumn{1}{c|}{$2.51\;s$} &
  \multicolumn{1}{c|}{$4.19\;s$} &
  \multicolumn{1}{c|}{$4.37\;s$} &
  $5.80\;s$ &
  \multicolumn{1}{c|}{$0.8\;ms$} &
  \multicolumn{1}{c|}{$0.6\;ms$} &
  \multicolumn{1}{c|}{$1.0\;ms$} &
  $1.1\;ms$ &
  \multicolumn{1}{c|}{$0.2\;ms$} &
  \multicolumn{1}{c|}{$1.0\;ms$} &
  \multicolumn{1}{c|}{$0.8\;ms$} &
  $0.8\;ms$ \\ \hline
\end{tabular}
\end{table*}






\begin{figure*}[h]
 \begin{center}
\includegraphics[scale =0.246]{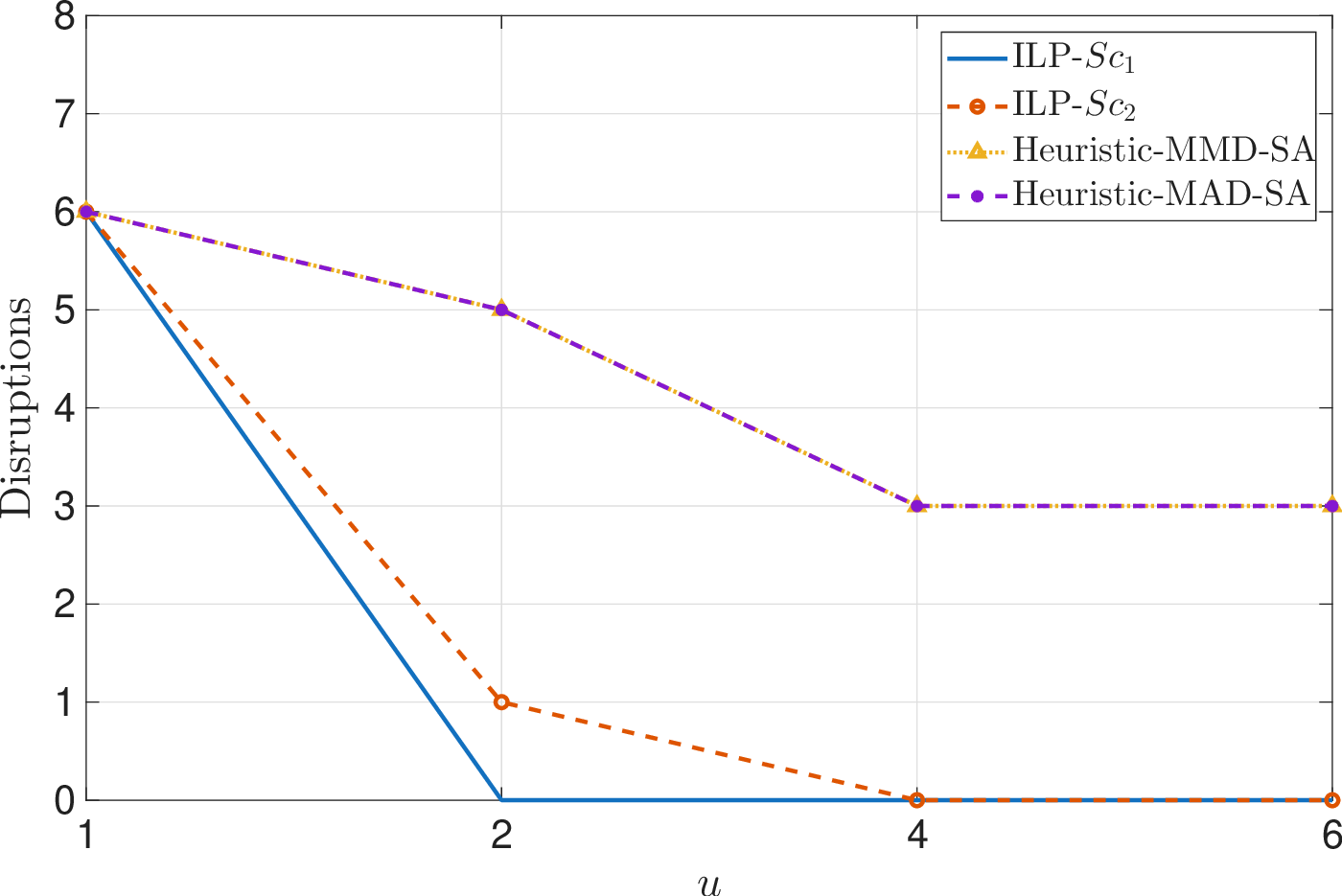}
\includegraphics[scale =0.246]{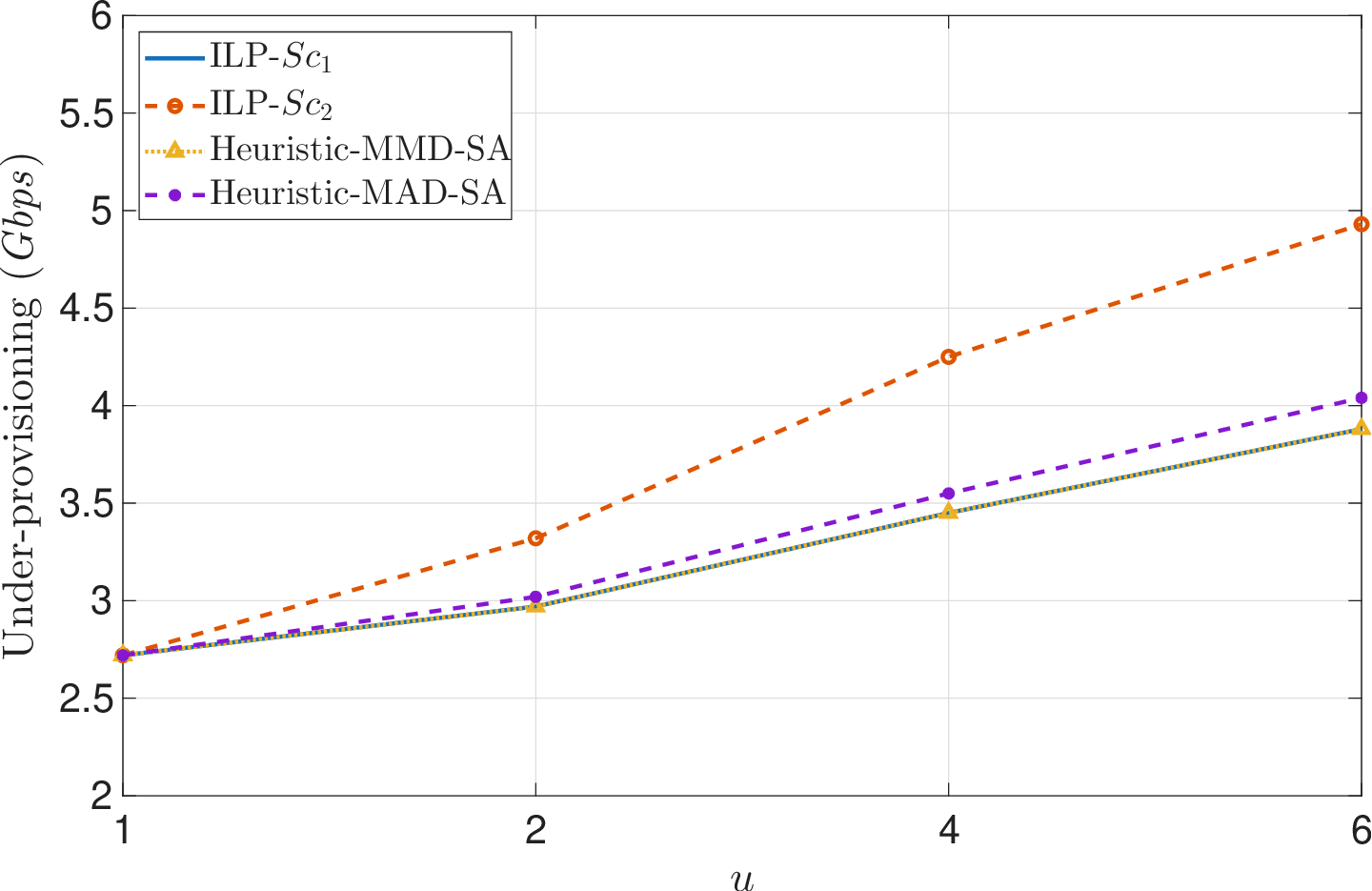}
\includegraphics[scale =0.246]{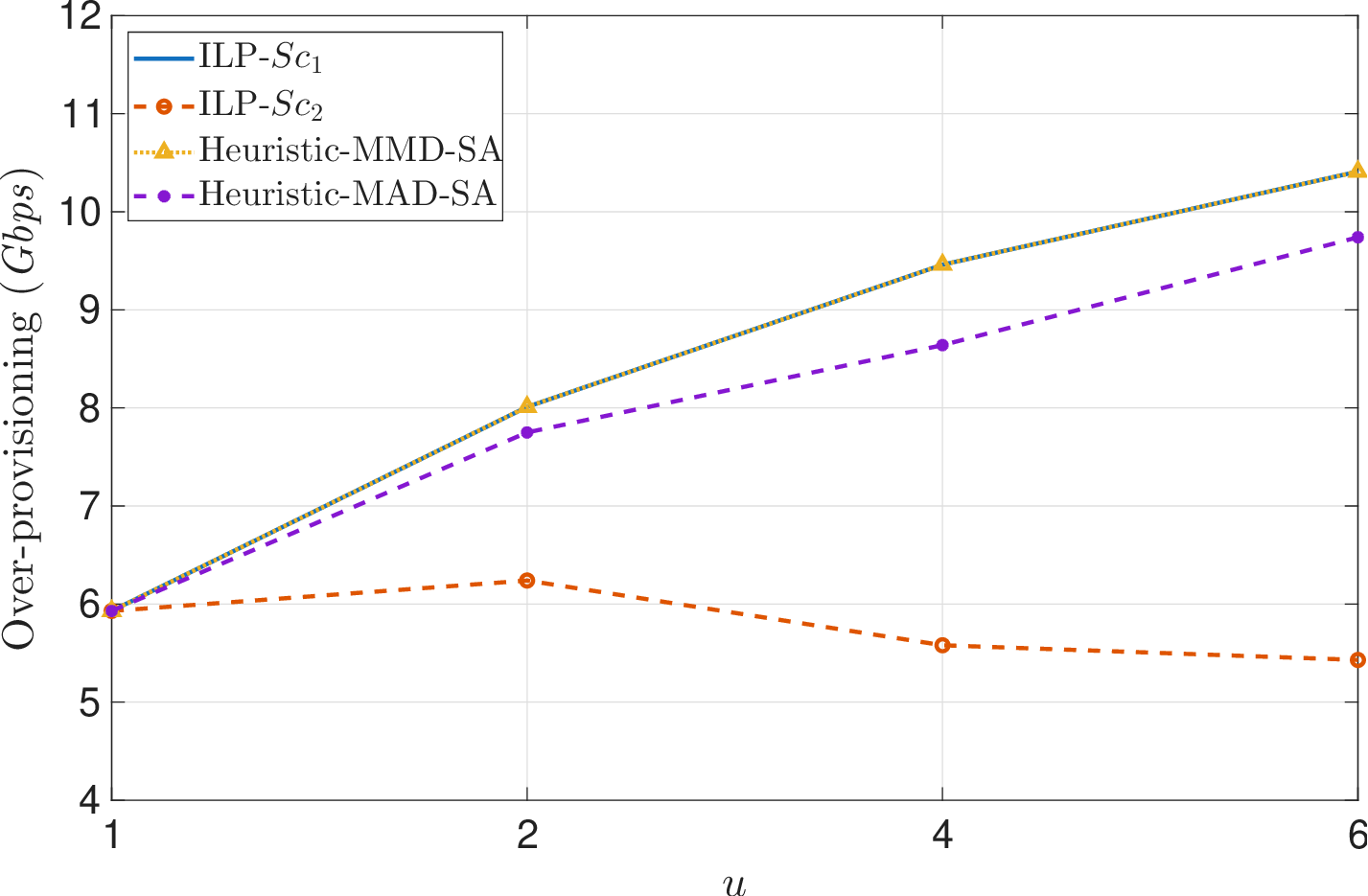}
 \vspace{-0.1in}
 \caption{Disruption, Over-provisioning, and Under-provisioning versus the number of multi-step ahead predictions ($u$).} \label{network_results}
\end{center}
\end{figure*}
\section{Performance Evaluation}~\label{evaluation}
Performance evaluation is performed on the Abilene network topology considering fiber links with capacity of $200$ FSs, each with a spacing of $12.5$ GHz, and a baud rate of $10.5$ Gbaud. Twelve ED-LSTM models are used, one for each $v$-$v'$ source-destination pair, with each destination node $v'$ randomly selected for each source node $v$. Both the predicted and true bit rate values are scaled (multiplied by $30$) to ensure reasonable spectrum demands given the capacity of the network. All schemes are assessed in the test datasets $D_v$ in terms of (i) number of blocked connections, (ii) number of disruptions, (iii) average under-provisioning measured in both Gbps and FSs, (iv) average over-provisioning measured in both Gbps and FSs, (v) average spectrum utilization, (vi) average maximum slot id ($F_{max}$), and (vii) average running times. It is noted that in the following results (presented at Table~\ref{ILP_heuristic_results}), under-provisioning and over-provisioning are computed by comparing the bit-rates selected against the true fluctuations in the test datasets. 

Specifically, for under- and over-provisioning in FS, true fluctuations are converted into FS (based on the modulation format supported by the selected path), and are then compared with the allocated FS of a connection. In this case, under-provisioning occurs when the number of true traffic fluctuation FSs are above the number of FSs allocated, and over-provisioning occurs when the true traffic fluctuation FSs are below the FSs allocated. In a similar manner, to account for the under- and over-provisioning Gbps rate, the true fluctuation rate is compared with the maximum and minimum rates satisfied by the allocated FSs (i.e., minimum: rate achieved if one less FS is allocated, and maximum: rate achieved by utilizing the allocated FSs). In this case, under-provisioning occurs when the true fluctuation rate is below the minimum rate, while over-provisioning occurs when the true fluctuation rate is above the maximum rate. It is noted that all under-provisioning and over-provisioning metrics are averaged over all planning intervals, connections, and fluctuations. Finally, a PC with an i7-12700K CPU, 64GB RAM, and a GTX 1660 Ti was used for the heuristic simulations and for training of all traffic prediction models, while each ILP simulation was performed on an HPC server with access to $10$ threads and $45$ GB RAM \cite{ucy_hpc}. For solving the ILP formulations, the Gurobi solver was used \cite{gurobi}. 

Regarding the ILP approach, to showcase the difference between the weights utilized for variables $Z$ and $V$, two scenarios are presented, where different combinations for $w_{2}$ and $w_{3}$ are selected, while weights $w_{1}$, $w_{4}$, and $w_{5}$ remain the same. These two scenarios are (i) $Sc_{1}$: the weight of $Z$ is greater than the weight of $V$ (\textit{focus on minimizing under-provisioning}), and (ii) $Sc_{2}$: the weight of $V$ is greater than the weight of $Z$ (\textit{focus on minimizing over-provisioning}). It is noted that several combinations of weights were empirically tested, and the set of combinations that provided the best results in terms of disruptions, over- and under-provisioning, and efficient spectrum utilization were selected, as shown in Table \ref{ILP_heuristic_results}. It is also important to note here again that both the ILP and heuristics consider only the predicted bit-rates at each period. Hence, any minimization of metrics associated with over- and under-provisioning highly depends on the quality of the predictions as well, since the true fluctuations are known only after the planning period has elapsed.

Table~\ref{ILP_heuristic_results} presents the results of the ILP (both $Sc_{1}$ and $Sc_{2}$) and both heuristic approaches (MMD-SA and MMA-SA) for different cases of the prediction horizon (i.e., $u=1,2,4,6$ prediction steps). Note that for $u$=$1$, all approaches are reduced to the conventional single-step-ahead prediction scheme, allocating to each connection request the bit-rate predicted for the next planning interval, which is used as a benchmark scenario. Additionally, Fig.~\ref{network_results} graphically presents key results of this work, namely disruptions, over-provisioning, and under-provisioning. While the figure illustrates the same results reported in Table~\ref{ILP_heuristic_results}, the graphical representation facilitates trend visualization and enables a clearer comparison of the behavior of the different approaches across prediction horizons.

As can be seen from the results, all approaches establish all connections through all periods considered (zero blocking). Starting from the benchmark case of $u=1$ (i.e., single-step-ahead predictions), all approaches achieve the same results in terms of over- and under-provisioning, as only a single bit-rate is predicted for each connection. Hence, as at each period the order of demands (i.e., ILP considers all demands together, while heuristics consider demands sequentially), and the resource provisioning (i.e., path selection and spectrum allocation) for each demand is different between the ILP and heuristic approaches, the ILP is able to utilize less spectrum slots to establish all connections compared to the solution provided by the heuristic approaches, at the expense of increasing $F_{max}$.   

Further, as the prediction horizon increases (i.e., number of prediction steps $u$), service disruptions decrease in almost all cases (for the heuristic approaches, disruptions remain the same for $u=4$ and $u=6$) through all proposed approaches (Fig.~\ref{network_results}). This is expected, and it occurs due to the allocation of larger spectrum sizes, as well as the fact that connections are subject to the event of a reconfiguration less frequently. As observed, the proposed ILP is able to achieve less or equal disruptions compared to the heuristics (that both obtain the same results with respect to disruptions), while in many cases, the ILPs produce solutions with zero disruptions. 


Utilizing larger prediction horizons leads to reduced disruptions, but also has trade-offs related to the quality of the allocation (in terms of over- and under-provisioning). Also, the bit-rate selection approach followed affects the quality of the solution (in terms of over- and under-provisioning). As can be seen from Fig.~\ref{network_results}, a general trend followed by the ILP for $Sc_{1}$ and both heuristic approaches is that over-provisioning tends to increase as the prediction horizon increases,  as higher bit-rates are likely to be opted to accommodate traffic for longer planning intervals, and hence a larger number of spectrum slots will likely be allocated. While in $S_{c_{1}}$ the ILP selects higher predicted bit-rates in an effort to reduce under-provisioning, this is not the case for scenario $Sc_{2}$, due to the increased weight of the term associated with minimizing over-provisioning ($w_{3}$) in the objective (as compared to $w_{2}$), which leads to selecting lower bit-rates (among the predicted ones) in each period. In $Sc_{2}$, the ILP achieves better solutions in terms of over-provisioning compared to both $Sc_{1}$ and the heuristic approaches, with up to $0.47$ slots (or up to $6.27$ Gbps) over-provisioned on average. Also, the MMD-MA heuristic approach achieves higher over-provisioning compared to MMA-SA, as it always considers the maximum predicted bit-rate, in an effort to reduce under-provisioning (similar to $Sc_{1}$).

Under-provisioning (Fig.~\ref{network_results}), on the other hand, is typically expected to decrease with longer prediction horizons under perfect traffic conditions (i.e., no prediction error). However, this is not the case in practice, and hence, increased prediction error ultimately degrades the performance of both over-provisioning and under-provisioning metrics. In addition, as also observed from the results, when the proposed approaches aim at reducing over-provisioning, the solutions obtained provide higher under-provisioning results, which also increases for larger prediction horizons. Nevertheless, it is noted that the $Sc_{1}$ scenario achieves reduced under-provisioning compared to the $Sc_{2}$ scenario. The exact same results are captured by MMD-SA as well (albeit with some additional slots utilized), with the MMA-SA approach achieving higher under-provisioning results (from $3.88$ Gbps to $4.04$ Gbps for $u=6$), at the benefit of slightly decreasing over-provisioning (from $10.41$ Gbps to $9.74$ Gbps for $u=6$). Further, scenario $Sc_{2}$ achieves more balanced results in terms of over- and under-provisioning by reducing over-provisioning by $4.98$ Gbps (from $10.41$ Gbps to $5.43$ Gbps for $u=6$), while increasing under-provisioning by $1.05$ Gbps (from $3.88$ Gbps to $4.93$ Gbps for $u=6$).

Overall, it can be seen from the results that the proposed ILP can provide different solutions in terms of over-provisioning and under-provisioning based on the objective weight configurations provided. Nevertheless, while the running time of the ILP is small for these instances (on the order of seconds as presented in Table~\ref{ILP_heuristic_results}), solving the problem for much larger (practical) instances would not be feasible, due to the nature of the RSA problem and the scalability of ILP-based approaches for such problems (the RSA problem is proven to be NP-complete \cite{chrisNPcomplete}). Hence, the more scalable heuristic approaches that achieve solutions close to the ILP in significantly less time, are more suitable solutions for larger network instances.

\section{Conclusion}\label{con}
This work demonstrates the importance of efficiently modeling and predicting multi-step ahead traffic for multi-period network optimization in optical networks. Specifically, an ML-aided technique, based on an ED-LSTM model, is employed to predict network traffic demand over long future horizons. Both, ILP and heuristic algorithms are designed and developed to effectively exploit the predictions. Specifically, it is shown that the proposed ILP and heuristic approaches, exploiting the multi-step-ahead predictions, are able to effectively establish connections in the network, by significantly reducing disruptions, compared to the case where single-step ahead predictions are utilized ($u=1$). Additionally, while the proposed ILP outperforms the heuristics in terms of the quality of the solutions obtained, both the MAD-SA and MMD-SA schemes enable network operators to effectively identify and handle the future traffic demand over a longer horizon (e.g., $2$ hours ahead) in a much more scalable way, while still achieving results close to the ILP. Future work entails the consideration of uncertainty to further improve spectrum savings.

\section*{Acknowledgments}
This work was supported by the European Union’s Horizon 2020 research and innovation programme under grant agreement No. 739551 (KIOS CoE - TEAMING) and 
by the Republic of Cyprus through the Deputy Ministry of Research, Innovation, and Digital Policy.


\bibliography{biblio}


\end{document}